\documentclass [aps,prb,preprint,superscriptaddress,showpacs]{revtex4}
\usepackage{braket}
\usepackage{color,soul} 
\usepackage{graphicx}
\usepackage{amsmath}
\usepackage{amsfonts}
\usepackage{xcolor}

\definecolor{LightGray}{gray}{0.8}
\definecolor{Orange}{rgb}{1.0, 0.31, 0.0}
\definecolor{Green}{rgb}{0.3, 1.0, 0.3}
\definecolor{Blue}{rgb}{0.75,0.75,1}

\begin{document}
\preprint{Accepted for publication in APS Open Science}
\title{Axion-mediated photon-to-photon transitions in high finesse dielectric resonators}
\author{Evangelos Almpanis} \email[e-mail address:~]{ealmpanis@gmail.com}
\affiliation{Physics Division, National Technical University of Athens, GR-157~80 Zografou Campus, Athens,
Greece}

\date{\today}

\begin{abstract}
Axions are hypothetical particles that could address both the strong charge-parity problem in quantum chromodynamics and the enigmatic nature of dark matter. However, if axions exist, their mass remains unknown, and they are expected to interact very weakly with the electromagnetic field, which explains why they have not been detected yet.  This study proposes a way to substantially augment the axion-photon interaction by confining the photons within high-quality-factor dielectric resonators, increasing their intensity and lifetime, and thus the possibility of interacting with axions in the background.
In view of this, we study resonant axion-mediated photonic transitions in millimeter-sized spherical dielectric resonators, based on fully analytical calculations to the first order in perturbation theory.
Such resonators exhibit high lifetime Mie resonances in the microwave part of the spectrum, with a separation that can be tailored with the radius of the sphere to match the expected axion frequency, allowing axion-mediated photonic transitions when particular selection rules are fulfilled.  
We predict experimentally accessible axion mass regimes where such triply resonant transitions can be realized with standard dielectric resonators. We propose an experiment for probing such interactions named DARK-ROSE.
\end{abstract}
\maketitle

\section{introduction}

One of the main mysteries in our current understanding of the physical world is dark matter, which is believed to constitute a significant portion of the total mass of the universe~[\onlinecite{rubin1970rotation,heinemeyer2008confronting,bertone2018history,jackson2023search,derocco2022dark}]. Despite its abundance and crucial role in explaining the structure and formation of the universe, dark matter remains undetected due to its weakly interacting nature. A potential candidate for this is the axion~[\onlinecite{choi2021recent}], an elementary particle proposed independently as a solution to the strong charge-parity ($\mathcal{CP}$) problem in quantum chromodynamics (QCD)~[\onlinecite{kim2010axions}]. Essentially, this problem revolves around the non-violation of $\mathcal{CP}$ symmetry by the strong nuclear force.
This is quantified by a parameter named $\theta$, which has been measured as zero ($\le 10^{-10}$) in all experiments~[\onlinecite{adams2022axion}],
suggesting a puzzling \emph{fine-tuning} problem. In order to solve this problem, in the
1970’s Roberto Peccei and Helen Quinn proposed a new field, known as the Peccei-Quinn field~[\onlinecite{peccei1977cp}],
from which the axion particle emerges~[\onlinecite{wilczek1978problem,weinberg1978new}]. Given that the existence of axions offers a promising
solution to both the enigmatic nature of dark matter and the strong $\mathcal{CP}$ problem, the research on them
is vitally important.

Motivated by the original proposal for axions discussed earlier, the concept was extended beyond the strict framework of QCD, leading to proposals that constitute a broader class of axion-like elementary particles~[\onlinecite{coriano2007stuckelberg,choi2021recent,kim2010axions,masso2005evading,marsh2017new,irastorza2018new,desjacques2018impact,arvanitaki2020large,mavromatos2022primordial,alexander2024field}] that serve more general theoretical and observational purposes. The common property of all axion and axion-like particles is that they are bosons that interact very weakly with the electromagnetic field through a mixing between the electric and the magnetic field. Interestingly, such a mixing of the electric and the magnetic field is possible (without the need for cosmic axions) in exotic solid-state materials, so-called magnetoelectric (ME) or Tellegen materials. This class of materials was first conceptually conceived by B. Tellegen~[\onlinecite{tellegen1948gyrator}], while independently, the ME effect (equivalent to Tellegen) was conceptualized by L. Landau~[\onlinecite{landau1884electrodynamics}] and predicted in $\mathrm{Cr}_2\mathrm{O}_3$ by I. Dzyaloshinskii~[\onlinecite{dzyaloshinskii1960magneto}], followed by experimental verifications~[\onlinecite{krichevtsov1993spontaneous}]. 
However, until now, the observed ME phenomena are negligible in most materials with the exception of certain topological insulators, also called
\emph{axion insulators}~[\onlinecite{wu2016quantized,dziom2017observation,varnava2018surfaces,seidov2023hybridization}], 
and more specifically, the antiferromagnetically-doped topological insulators, where
\emph{axion quasiparticles} can arise due to the intrinsic magnetic properties and the unique structure of the
material~[\onlinecite{marsh2019proposal,schutte2021axion,esposito2023optimal,catinari2025hunting}].

Now, a possible enhanced coupling between axion quasiparticles (solid-state axions) and cosmic axions~[\onlinecite{marsh2019proposal,schutte2021axion}] has generated proposals for a new generation of cosmic axion detectors named TOORAD (TOpOlogical Resonant Axion Detection)~[\onlinecite{semertzidis2022axion}]. We note here that searching for cosmic axions is like tuning a radio: it needs a specific frequency but we don’t know what that frequency is (where the role of the frequency is played by the mass of the axion). In this respect, 
the TOORAD experiment will search for cosmic axions with a mass of $0.7$ to $3.5$ meV$/c^2$, it complements, rather than replaces, more established experimental avenues such as haloscopes and helioscopes. Haloscopes, like ADMX~[\onlinecite{nitta2023search,mcallister2024tunable,carosi2025search}], are designed to detect axions from the Galactic dark matter halo (masses~$\sim\mu$eV$/c^2$) by converting them into photons inside a microwave cavity tuned to the axion mass. More recently, dielectric Haloscopes~[\onlinecite{millar2017dielectric,caldwell2017dielectric}] and plasma Haloscopes~[\onlinecite{terccas2018axion,lawson2019tunable,millar2023searching}] have also been proposed. 
In contrast, helioscopes such as CAST~[\onlinecite{zioutas2005first,arik2009probing}] and the next-generation IAXO~[\onlinecite{armengaud2014conceptual,armengaud2019physics}] aim to detect axions produced in the Sun via their conversion into x-rays in a strong magnetic field. Together, these approaches, along with others, cover complementary regions of the axion parameter space.

If axions (pseudoscalar bosons with spin $\mathcal{S}=0$) exist, they will interact with the photon (vector boson with spin $\mathcal{S}=1$) in a very specific way. A photon $\gamma_\mathrm{i}$ could absorb an axion $\alpha$ to produce a photon $\gamma_\mathrm{f}$ with increased frequency, i.e., $\gamma_{\mathrm{i}} +\alpha\rightarrow \gamma_{\mathrm{f}} $. Alternatively, a photon $\gamma_\mathrm{i}$ could produce an axion $\alpha$ and a photon $\gamma_\mathrm{f}$ with a lower frequency ($\gamma_{\mathrm{i}} \rightarrow \gamma_{\mathrm{f}}+\alpha $). 
Such processes involve two photons and one axion~[\onlinecite{galanti2022axion}] are obviously allowed and can be viewed as axion-mediated transitions from photon $\gamma_\mathrm{i}$ to photon $\gamma_\mathrm{f}$. We note here that for photons with energies in the $\sim\mu$eV scale that we will consider in this work, the momentum is on the order of $\sim\mu$eV/$c$, while the momentum of nonrelativistic dark matter axions (like the ones investigated with Haloscopes~[\onlinecite{millar2017dielectric}]) with mass $\sim\mu$eV$/c^2$ is expected to be on the order of $\sim$neV/$c$, which is negligibly small compared to that of the photons. In view of this, the axion acts as an energy reservoir, enabling photonic transitions without significant momentum mismatch~[\onlinecite{murgui2024axion}].

Since the axion-photon-photon coupling constant $g_{\alpha\gamma\gamma}$ is very small, here we propose a strategy to substantially enhance the photon intensity and lifetime inside a high quality-factor ($Q$) dielectric resonator, increasing in this way the probability for axion-photon interactions when axions with specific mass exist in the background, and particular selection rules are being respected. A conceptually related approach employing (super)conducting cylindrical cavities was discussed in Refs.~\mbox{[\onlinecite{berlin2020axion,thomson2021upconversion,thomson2023searching}]}; however, here we extend the concept to all-dielectric spherical resonators, providing fully analytical calculations to first order in perturbation theory and deriving the corresponding selection rule based on group theory. At first, in Sec.~\ref{theory} we define the distinct multipolar eigenstates of the electromagnetic (EM) field inside a spherical dielectric resonator. In Sec.~\ref{axion-photon} we come up with the axion-photon interaction expressed through a new term in the Maxwell Lagrangian. In Sec.~\ref{transitions} we set up the resonant axion-mediated photonic transitions evaluated to first order in perturbation theory and establish the selection rules that govern such transitions. This approach is sufficient for our purposes, while at the same time allowing for a deeper insight into the underlying physics. In Sec.~\ref{results}, which is devoted to the discussion of our results, we provide our fully analytical calculations for the enhancement of the axion-mediated photonic transition rates in the presence of an optical resonator and give specific results. The last section summarizes the main findings of the paper.

\section{Photonic resonance frequencies of a spherical Mie resonator}
\label{theory}

\begin{figure}[h]
    \centering
    \includegraphics[width=0.8\linewidth]{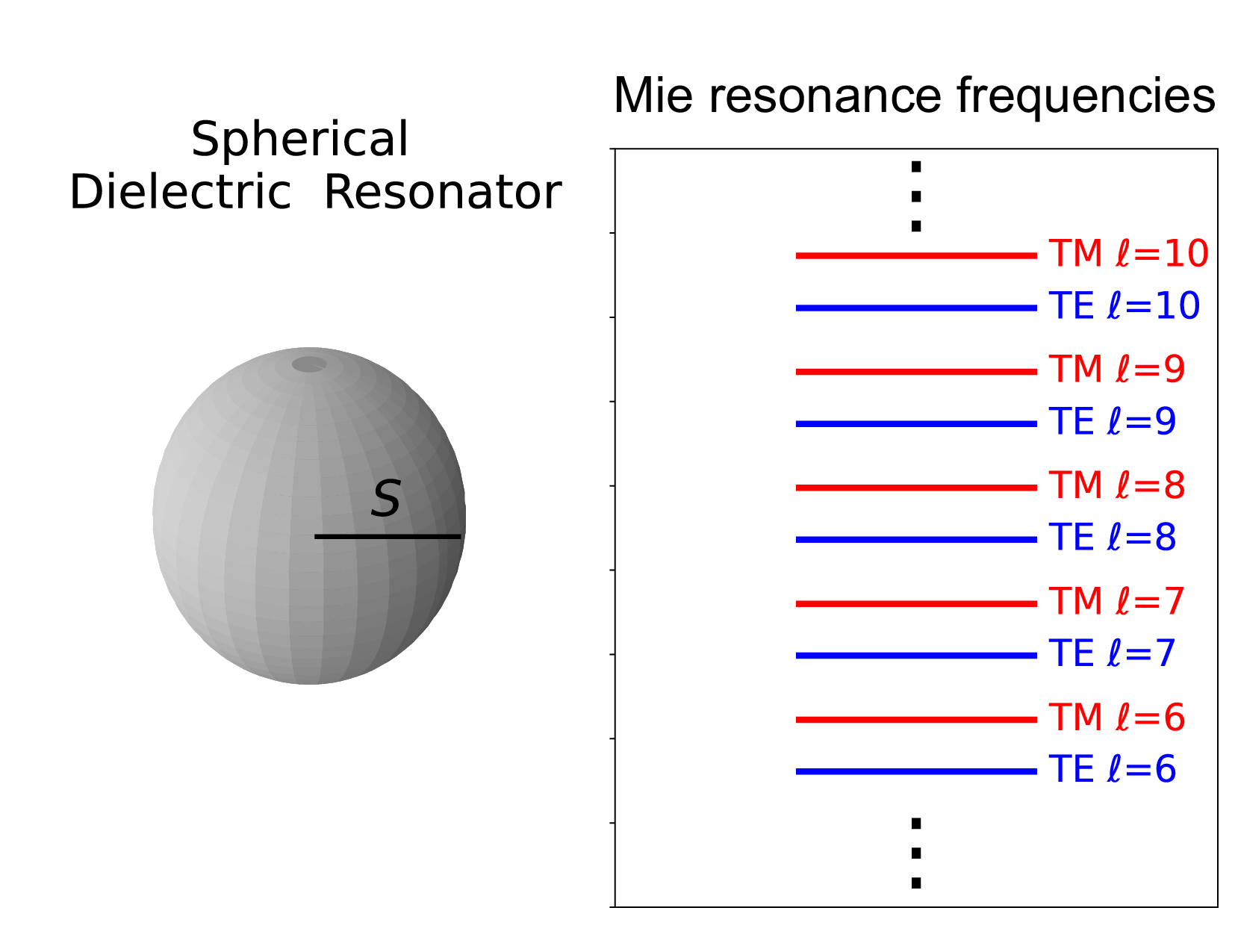}
    \caption{Discrete photonic Mie resonance frequencies for a spherical dielectric resonator of radius $S$ corresponding to fundamental ($\nu=1$) TE (blue) and TM (red) modes of angular momentum $\ell$. The dots indicate the presence of additional modes beyond the displayed range. 
    }
    \label{fig_01}
\end{figure}

We will consider a homogeneous and isotropic dielectric sphere (relative permittivity $\epsilon$, relative permeability $\mu$) of radius $S$ in air (free space permittivity $\epsilon_0$, free space permeability $\mu_0$). Scattering of light by such a sphere can be solved analytically in the framework of Mie theory~[\onlinecite{mie1908beitrage,bohren2008absorption,jackson2021classical}], assuming a spherical expansion of both incoming and outgoing fields. By doing so, multiple peaks appear in the corresponding scattering cross section spectrum, the so-called optical Mie resonances. Each Mie mode is characterized by the polarization type: magnetic type (TE) or electric type (TM), the angular momentum index $\ell$, and the magnetic number $m$. We note here that due to spherical symmetry, these modes are degenerate ($2\ell+1$ degeneracy). Although these modes have a finite lifetime, when the refractive index of the scatterer is high and the losses are small, the lifetime is long, leading to measured quality factors on the order of $Q \sim 10^{1}-10^3$ for materials such as silicon[\mbox{\onlinecite{fenollosa2023silicon}}], titania[\mbox{\onlinecite{mitrofanov2014near,lukyanchuk2021colossal}}], alumina[\mbox{\onlinecite{dey2021low,subrahmannian2022study}}] or other novel ceramics[\mbox{\onlinecite{galante2025observation}}] for small multipole indices  ($\ell\lesssim5$), depending on the particle size and the frequency band. The quality factor of the resonances scales with the $\ell$ index[\mbox{\onlinecite{lam1992explicit}}], while for high values of $\ell$ (whispering gallery modes) the measured quality factors are at the order of $\sim10^5$ to $10^{10}$~\mbox{[\onlinecite{gorodetsky1996ultimate,yu2012spherical,rueda2016efficient,vogt2018ultra,lambert2020coherent,ilchenko2013whispering}]}. We note here that these quality factors correspond to room temperature. Throughout this work, we restrict our analysis to fundamental modes with radial index $\nu=1$.
In Fig.~\ref{fig_01}, we schematically show the resonance frequencies of the different spectrally separated Mie modes of a spherical dielectric resonator.   
The electric and magnetic fields, inside the sphere, of a TE mode with resonance frequency $\omega_1$ are given by the expressions~[\onlinecite{stefanou1998heterostructures,almpanis2020spherical,almpanis2021spherical}]

\begin{eqnarray}
    {\bf E}_{M\ell m}({\bf r};\omega_1) = a_{M\ell m} j_{\ell}(q_1r) {\bf X}_{\ell m}(\hat{\bf r}) \nonumber \\
    {\bf B}_{M\ell m}({\bf r};\omega_1) =  -i\frac{\sqrt{\epsilon\mu}}{q_1 c} a_{M\ell m} \nabla \times \big( j_{\ell}(q_1r) {\bf X}_{\ell m}(\hat{\bf r}) \big),
    \label{TE}
\end{eqnarray}
while those of a TM mode with resonance frequency $\omega_2$ are given by the expressions

\begin{eqnarray}
    {\bf E}_{E\ell m}({\bf r};\omega_2) = \frac{i}{q_2 }a_{E\ell m} \nabla \times \big( j_{\ell}(q_2r) {\bf X}_{\ell m}(\hat{\bf r}) \big)           \nonumber \\
    {\bf B}_{E\ell m}({\bf r};\omega_2) = \frac{\sqrt{\epsilon\mu}}{c}  a_{E\ell m} j_{\ell}(q_2r) {\bf X}_{\ell m}(\hat{\bf r}),
    \label{TM}
\end{eqnarray}
where $q_{1,2}=\omega_{1,2}\sqrt{\epsilon \mu}/c$, $c=1/\sqrt{\epsilon_0 \mu_0}$ being the velocity of
light in vacuum; $j_{\ell}$ are the spherical Bessel functions
which are finite everywhere; and $\mathbf{X} _{\ell
m}(\hat{\mathbf{r}})$ are the vector spherical harmonics, $\sqrt{\ell(\ell+1)}\mathbf{X} _{\ell
m}(\hat{\mathbf{r}}) \equiv -i{\bf{r}} \times \nabla Y_{\ell m}(\hat{\mathbf{r}})$. The expansion coefficients
$a_{P\ell m}$, $P=E,M$, have units of electric field (V/m).

The symmetry group of a spherical particle, which involves both proper and improper rotations, is the $O(3)$ Lie group. The character table of the $O(3)$ group is shown in Table~\ref{tab:LieGroup}. The subscripts $\mathrm{g}$ (gerade) and $\mathrm{u}$ (ungerade) denote the parity of the multipole, i.e., $\mathrm{g}$ for even parity and $\mathrm{u}$ for odd parity. We note here that if we assign the index $P=1$ for magnetic multipoles (TE) and $P=2$ for electric multipoles (TM), then the sum $P+\ell$ defines the parity of the Mie mode (even or odd)~[\onlinecite{gantzounis2009plasmon,almpanis2021spherical}].    

\begin{table}[!htb]
   \centering
        \begin{tabular}{c| c c c c}     \hline
            $O(3)$ &    $R_{\theta}$  &  $\mathcal{I}R_{\theta}$~  \\  \hline \hline
            $D_{\mathrm{g}}^{(\ell=0)}$&  1  & 1 \\    \hline
            $D_{\mathrm{u}}^{(\ell=0)}$&  1 &  $-1$   \\    \hline
            $D_{\mathrm{g}}^{(\ell)}$&  $\frac{\sin{(2\ell+1)\theta/2}}{\sin{\theta/2}}$ &  $\frac{\sin{(2\ell+1)\theta/2}}{\sin{\theta/2}}$  \\    \hline
            $D_{\mathrm{u}}^{(\ell)}$&  $\frac{\sin{(2\ell+1)\theta/2}}{\sin{\theta/2}}$ &  $-\frac{\sin{(2\ell+1)\theta/2}}{\sin{\theta/2}}$  \\    \hline
        \end{tabular}
    \caption{Character table of the $O(3)$ Lie group ($\ell=1,2,3,\ldots$). $R_{\theta}$ are rotation operations through an angle $\theta$ ($\theta \in [0, \pi]$) about an axis, and $\mathcal{I}$ is the inversion operation.}\label{tab:LieGroup}
\end{table}

\section{Axion-photon interaction}
\label{axion-photon}

The interaction between axion and photon is described by an additional term in the Maxwell Lagrangian~[\onlinecite{roising2021axion,adams2022axion,galanti2022axion,berlin2024absorption,sikivie1983experimental,kim2019effective,sikivie2024axion}] (Lagrangian density, units eV/m$^3$ )

\begin{equation}
    {\mathcal{L}}_{\alpha\gamma\gamma} = -g_{\alpha\gamma\gamma} \alpha({\bf{r}},t) {\bf E} \cdot {\bf B}, 
\end{equation}
where $\alpha({\bf{r}},t)$ is the 
axion field in units of eV, $g_{\alpha\gamma\gamma}=\Tilde{g}_{\alpha\gamma\gamma}/\mu_0 c$, where $\Tilde{g}_{\alpha\gamma\gamma}$ the axion-photon-photon coupling constant expressed in units eV$^{-1}$. We note here that, although the factor $1/\mu_0 c$ is taken as unity in natural units, it must be explicitly included when working in the SI system~\mbox{[\onlinecite{visinelli,terccas2018axion,kim2019effective,brevik2021axion}]}, as we do throughout this paper. In the presence of this term, the Maxwell equations, in a linear medium ($\epsilon, \mu$) without conventional~[\onlinecite{griffiths2023introduction}] free sources (the axionic terms now play the role of sources), can be recast as follows~[\onlinecite{adams2022axion,wilczek1987two,sikivie1983experimental,kim2019effective}]

\begin{eqnarray}
  \nabla \cdot (\epsilon_0\epsilon\mathbf{E}) = g_{\alpha\gamma\gamma} {\bf B} \cdot \nabla\alpha({\bf{r}},t), ~~~~ \nabla \cdot \mathbf{B} = 0 ~~~~~~~~~~~~~~~~~~~~~~~~~\nonumber \\
\nabla \times \mathbf{E} = -\frac{\partial \mathbf{B}}{\partial t},~~~~~  \nabla \times (\frac{1}{\mu_0\mu}\mathbf{B}) = \frac{\partial(\epsilon_0 \epsilon \mathbf{E})}{\partial t} +g_{\alpha\gamma\gamma}\big({\bf E}\times\nabla\alpha({\bf{r}},t) -\frac{\partial\alpha({\bf{r}},t)}{\partial t}\big)
\end{eqnarray}
whereby by setting the following constitutive relations

\begin{eqnarray}
    {\bf D} = \epsilon_0\epsilon{\bf E} - g_{\alpha\gamma\gamma} \alpha({\bf{r}},t) {\bf B}    
    \label{Dfield}\\
    {\bf H} = \frac{1}{\mu_0\mu}{\bf B} + g_{\alpha\gamma\gamma} \alpha({\bf{r}},t) {\bf E}
    \label{Hfield}
\end{eqnarray}
we obtain the following set of equations

\begin{eqnarray}
\nabla \cdot \mathbf{D} = 0, ~~~~~~~ \nabla \cdot \mathbf{B} = 0 ~~~~~ \nonumber \\
\nabla \times \mathbf{E} = -\frac{\partial \mathbf{B}}{\partial t}, ~~~~ \nabla \times \mathbf{H} = \frac{\partial \mathbf{D}}{\partial t} 
\label{eqmatter}
\end{eqnarray}

Equations~(\ref{eqmatter}) are identical to the common Maxwell equations in matter~[\onlinecite{griffiths2023introduction}] (without free sources). This is analogous to light traveling in an \emph{axionic} medium~[\onlinecite{wilczek1987two,seidov2023hybridization}] that mixes the ${\bf E}$ and ${\bf B}$ fields.

\section{Axion-mediated photonic transitions}
\label{transitions}

In the case of cold dark matter, the axion can be considered as a coherent pseudo-scalar classical field~[\onlinecite{marsh2019proposal,chigusa2020detecting,schutte2021axion,semertzidis2022axion}] 

\begin{equation}
    \alpha({\bf{r}},t) = A({\bf{r}}) \cos{\Omega_{\alpha} t},
\end{equation}
where $\Omega_{\alpha} = m_{\alpha}c^2/\hbar$ the axion angular frequency and $m_{\alpha}$ is the axion mass. In this treatment, we approximate the galactic axion field as monochromatic, since its expected quality factor is very large, namely $Q_{\alpha} \sim 10^6$ \mbox{[\onlinecite{turner1990periodic,graham2015experimental,quiskamp2022direct}]}.
Since the axionic field varies in time with angular frequency $\Omega_{\alpha}$ it can induce inelastically scattered light beams with angular frequencies $\omega-\Omega_{\alpha}$ (Stokes) and $\omega+\Omega_{\alpha}$ (anti-Stokes), as in typical Brillouin (inelastic) scattering of light by matter excitations~[\onlinecite{wolff2021brillouin,gantzounis2011nonlinear,almpanis2018dielectric,almpanis2020spherical,stefanou2021light,panagiotidis2022inelastic}]. In the special case where $\omega$ is a resonance frequency of the photonic resonator and $\omega\pm\Omega_{\alpha}$ is also a (discrete) resonance frequency of the resonator, the so-called (triply) resonant transitions can occur~[\onlinecite{haigh2016triple,zhang2016optomagnonic,osada2016cavity,panagiotidis2023optical,sikivie2014axion}], provided that this is allowed by symmetry (as we shall see later on).  We now proceed by modeling this interaction. 
The constitutive relations~(\ref{Dfield}) and ~(\ref{Hfield}) can be cast as a linear system in matrix form as follows

\begin{eqnarray}
\left(
\begin{array}{c}
\mathbf{D} \\
\mathbf{H}
\end{array}
\right)
=
\left(
\begin{array}{cc}
\epsilon_0 \epsilon &- g_{\alpha\gamma\gamma} \alpha({\bf{r}},t) \\
 g_{\alpha\gamma\gamma} \alpha({\bf{r}},t) & \dfrac{1}{\mu_0 \mu}
\end{array}
\right)
\left(
\begin{array}{c}
\mathbf{E} \\
\mathbf{B}
\end{array}
\right)
\end{eqnarray}
where the $2\times2$ matrix consists of one static part and one time-dependent part, as follows

\begin{eqnarray}
\left(
\begin{array}{cc}
\epsilon_0 \epsilon & -g_{\alpha\gamma\gamma} \alpha({\bf{r}},t) \\
 g_{\alpha\gamma\gamma} \alpha({\bf{r}},t) & {1}/{\mu_0 \mu}
\end{array}
\right)
=
\left(
\begin{array}{cc}
\epsilon_0 \epsilon & 0 \\
0 & {1}/{\mu_0 \mu}
\end{array}
\right)
+
\left(
\begin{array}{cc}
0 & -g_{\alpha\gamma\gamma} \alpha({\bf{r}},t) \\
g_{\alpha\gamma\gamma} \alpha({\bf{r}},t) & 0
\end{array}
\right)
\end{eqnarray}

By assuming $\exp{(-i\omega t)}$ time dependence of the fields, we write the corresponding classical Hamiltonian (energy) density in the following form

\begin{eqnarray}
\mathcal{H}  
&=& 
\frac{1}{4}(\mathbf{E}^*\cdot \mathbf{D}+\mathbf{B}^* \cdot \mathbf{H})
=\frac{1}{4}
\left(
\begin{array}{cc}
\mathbf{E}^* & \mathbf{B}^*
\end{array}
\right)
\left(
\begin{array}{c}
\mathbf{D} \\
\mathbf{H}
\end{array}
\right) \nonumber \\
&=& \frac{1}{4}
\left(
\begin{array}{cc}
\mathbf{E}^* & \mathbf{B}^*
\end{array}
\right)
\left(
\begin{array}{cc}
\epsilon_0 \epsilon & 0 \\
0 & {1}/{\mu_0 \mu}
\end{array}
\right)
\left(
\begin{array}{c}
\mathbf{E} \\
\mathbf{B}
\end{array}
\right)  +
\left(
\begin{array}{cc}
\mathbf{E}^* & \mathbf{B}^*
\end{array}
\right)
\frac{g_{\alpha\gamma\gamma} \alpha({\bf{r}},t)}{4}
\left(
\begin{array}{cc}
0 & -1 \\
1 & 0
\end{array}
\right)
\left(
\begin{array}{c}
\mathbf{E} \\
\mathbf{B}
\end{array}
\right) \nonumber \\
&& \nonumber \\
&\equiv& \mathcal{H}_0 + \delta \mathcal{H}(t),
\end{eqnarray}
with the star denoting complex conjugation.
The $\mathcal{H}_0$ is a typical Hamiltonian density for an electromagnetic wave traveling in a linear medium~[\onlinecite{griffiths2023introduction}] having relative permittivity $\epsilon$ and relative permeability $\mu$ (unperturbed Hamiltonian), while the term $\delta\mathcal{H}(t)$ that contains all the \emph{axionic} information can be considered as a small time-dependent perturbation. The axion-mediated photonic transitions arise from this dynamical perturbation of the system. 

Since the axion-photon coupling is very weak, we can apply the Born approximation up to the first order to calculate the transition amplitudes. In view of this, the useful information is provided by the overlap integral $G$ (transition matrix element), where the perturbation matrix
$\delta \hat{V}(t)$
is sandwiched between the final and initial states, i.e, $G=\bra{\mathrm{f}}\delta \hat{V}(t)\ket{\mathrm{i}}$, where 
\begin{equation}
    \ket{\mathrm{i}} = 
    \left(
    \begin{array}{c}
        \mathbf{E}_{\mathrm{i}} \\
        \mathbf{B}_{\mathrm{i}}
    \end{array}
    \right)
    e^{-i\omega_{\mathrm{i}} t}
\end{equation}

\begin{equation}
    \bra{\mathrm{f}} = 
    \left(
    \begin{array}{cc}
        \mathbf{E}^*_{\mathrm{f}} & \mathbf{B}^*_{\mathrm{f}}
    \end{array}
    \right)
    e^{i\omega_{\mathrm{f}} t},
\end{equation}
and
\begin{equation}
    \delta \hat{V}(t) = 
    g_{\alpha\gamma\gamma} \frac{A(\mathbf{r}) \cos(\Omega_{\alpha} t)}{4}
    \left(
    \begin{array}{cc}
        0 & -1 \\
        1 & 0
    \end{array}
    \right)
    = 
    \frac{g_{\alpha\gamma\gamma} A(\mathbf{r})}{2}
    \left(
    \begin{array}{cc}
        0 & -1 \\
        1 & 0
    \end{array}
    \right)
    \left(e^{-i\Omega_{\alpha} t} + e^{i\Omega_{\alpha} t} \right).
\end{equation}

The overlap integral (transition matrix element) reads
\begin{equation}
G=\pi[\delta(\omega_\mathrm{i}-\omega_\mathrm{f}+\Omega_{\alpha})+\delta(\omega_\mathrm{i}-\omega_\mathrm{f}-\Omega_{\alpha})]g\;, \label{eq:scatelement}
\end{equation}
where 
\begin{equation}
g= g_{\alpha\gamma\gamma}   \int_V{\rm d}^{3}r A({\bf{r}})\big(\mathbf{B}^\ast_{\mathrm{f}}(\mathbf{r})\cdot\mathbf{E}_{\mathrm{i}}(\mathbf{r})-\mathbf{E}^\ast_{\mathrm{f}}(\mathbf{r})\cdot\mathbf{B}_{\mathrm{i}}(\mathbf{r}) \big). \label{g}
\end{equation}

The $\delta$ functions in Eq.~\ref{eq:scatelement} express energy conservation in the optical transitions that involve absorption and emission of one
axion by a photon. The volume $V$ is taken as the volume of the sphere since the electromagnetic field decays rapidly outside of the resonator. 

Now, it is straightforward to show that $\delta \hat{V}$ remains invariant under proper rotations $R_{\theta}$, while changes sign under inversions ($\mathcal{I}$) and improper rotations ($\mathcal{I}R_{\theta}$), since $A({\bf{r}}) $ is a pseudoscalar, i.e., $A(\mathbf{r}) \xrightarrow{E,~R_{\theta}} 1\cdot A(\mathbf{r})$ and $A(\mathbf{r}) \xrightarrow{\mathcal{I},~\mathcal{I}R_{\theta}} -1\cdot A(\mathbf{r})$. 
Therefore, $\delta \hat{V}$ is an irreducible tensor operator, which has the symmetry of the  $D_{\mathrm{u}}^{(\ell=0)}$ irreducible representation of the Lie group $O(3)$. This means that  $\delta \hat{V}$ operating on an eigenvector of the $P\ell$ irreducible subspace, transforms according to the relevant direct product representation

\begin{eqnarray}
    \label{eq:decomp}
    D_{\mathrm{u}}^{(\ell=0)}\otimes D_{\mathrm{g}}^{(\ell)} &=& D_{\mathrm{u}}^{(\ell)}\;, \\
    \nonumber
    D_{\mathrm{u}}^{(\ell=0)}\otimes D_{\mathrm{u}}^{(\ell)} &=& D_{\mathrm{g}}^{(\ell)} \;\;
\end{eqnarray}

\begin{figure}[h]
    \centering
    \includegraphics[width=0.6\linewidth]{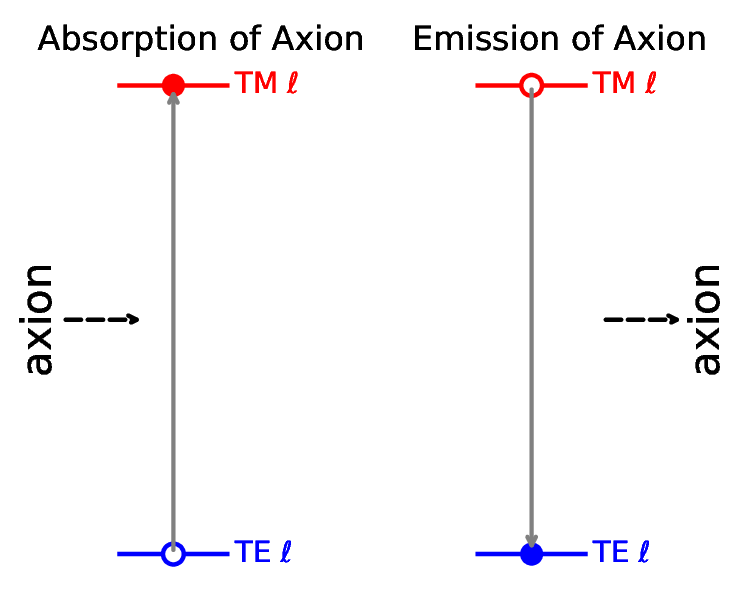}
    \caption{
    Schematic representation of axion-induced photonic transitions between electromagnetic modes in a spherical resonator. Left: An axion is absorbed, enabling a transition from a transverse electric (TE) mode to a transverse magnetic (TM) mode with the same angular momentum $\ell$.
    Right: The reverse process, where a transition from TM to TE mode is accompanied by the emission of an axion. The initial and final photonic states are marked with open and solid circles, respectively, while axions are represented by dashed arrows.
     }
    \label{fig_02}
\end{figure}

These considerations lead to a straightforward selection rule for the photonic transitions described above: the photonic mode must undergo a change in parity, while the angular momentum index $\ell$ remains conserved, for axion absorption (anti-Stokes) and axion emission (Stokes) processes. 
In Fig.~\ref{fig_02} we show a schematic representation of axion-mediated photonic transitions with respect to the mentioned selection rule, provided that the frequency difference of the optical modes is $\Delta f = \Omega_{\alpha}/2\pi$, where $\Omega_{\alpha}/2\pi$ is the frequency of the axion. In the left-hand panel we show the anti-Stokes process, where an axion is absorbed by the photonic transition from the TE$_{\ell}$ mode with lower frequency to the TM$_{\ell}$ mode with higher frequency, while in the right-hand panel we show the inverse process (Stokes) where an axion is emitted by the photonic transition from the TM$_{\ell}$ mode with higher energy to the TE$_{\ell}$ mode with lower energy.  
We also note that the same selection rule is deduced by explicitly (analytically) doing the algebra for the $\int {\bf B}^*_{\mathrm{f};P\ell m} \cdot {\bf E}_{\mathrm{i};P'\ell' m'} $ and  $\int {\bf E}^*_{\mathrm{f};P\ell m} \cdot {\bf B}_{\mathrm{i};P'\ell' m'} $ products and making use of the angular integration properties of the vector spherical harmonics. {We note here that these integrals also produce a factor $\delta_{mm'}$, which implies that that the initial and final photon modes must have the same azimuthal index $m$. In scatterers of perfectly spherical shape this index is degenerate, whereas in realistic experiments this degeneracy is typically lifted by small asymmetries such as weak deformations from the spherical shape, localized coupling elements (loop or prism couplers) positioned at fixed azimuthal angles, or controlled surface defects, among others.}

\section{Results and Discussion}
\label{results}

We now assume a spherical lossless high-index dielectric particle of radius $S$ in air. The relative dielectric permittivity of the sphere is $\epsilon=12$, while its relative magnetic permeability is $\mu=1$. We will present our results in scaled units of the radius $S$, nevertheless the radius $S$ can vary from microns to centimeters, depending on the targeted frequency range (we will give specific examples later). As discussed earlier, such particles support long-lifetime, spectrally separated, Mie modes of the EM field, characterized by the indices $P$ and $\ell$. For spherical particles, the resonant frequencies can be calculated analytically using the $T$-matrix formulation, which is evaluated on the particle's boundaries[\mbox{\onlinecite{mishchenko2002scattering}}]. Consequently, the refractive index of the surrounding medium shifts the resonant frequencies accordingly. Since in such particles the TE$_{\ell}$ mode has always lower energy than the corresponding TM$_{\ell}$ mode~[\onlinecite{bohren2008absorption}] (see Fig.~\ref{fig_01}), as the initial state we set the $\ell$ multipole of magnetic type (TE) and as the final state we take the $\ell$ multipole of electric type (TM), for the axion absorption process, to comply with the selection rule. The frequency difference between the initial and final states must be adjusted to the axion frequency, $f_{\mathrm{f}}-f_{\mathrm{i}}=\Omega_{\alpha}/2\pi$, to obtain the resonant transition. We note here that the reverse process (axion emission) is also possible, as shown in the right-hand panel of Fig.~\ref{fig_02}, and the transition matrix element is the same, so we will not discuss it for time saving.    

\begin{figure}[h]
    \centering
    \includegraphics[width=0.7\linewidth]{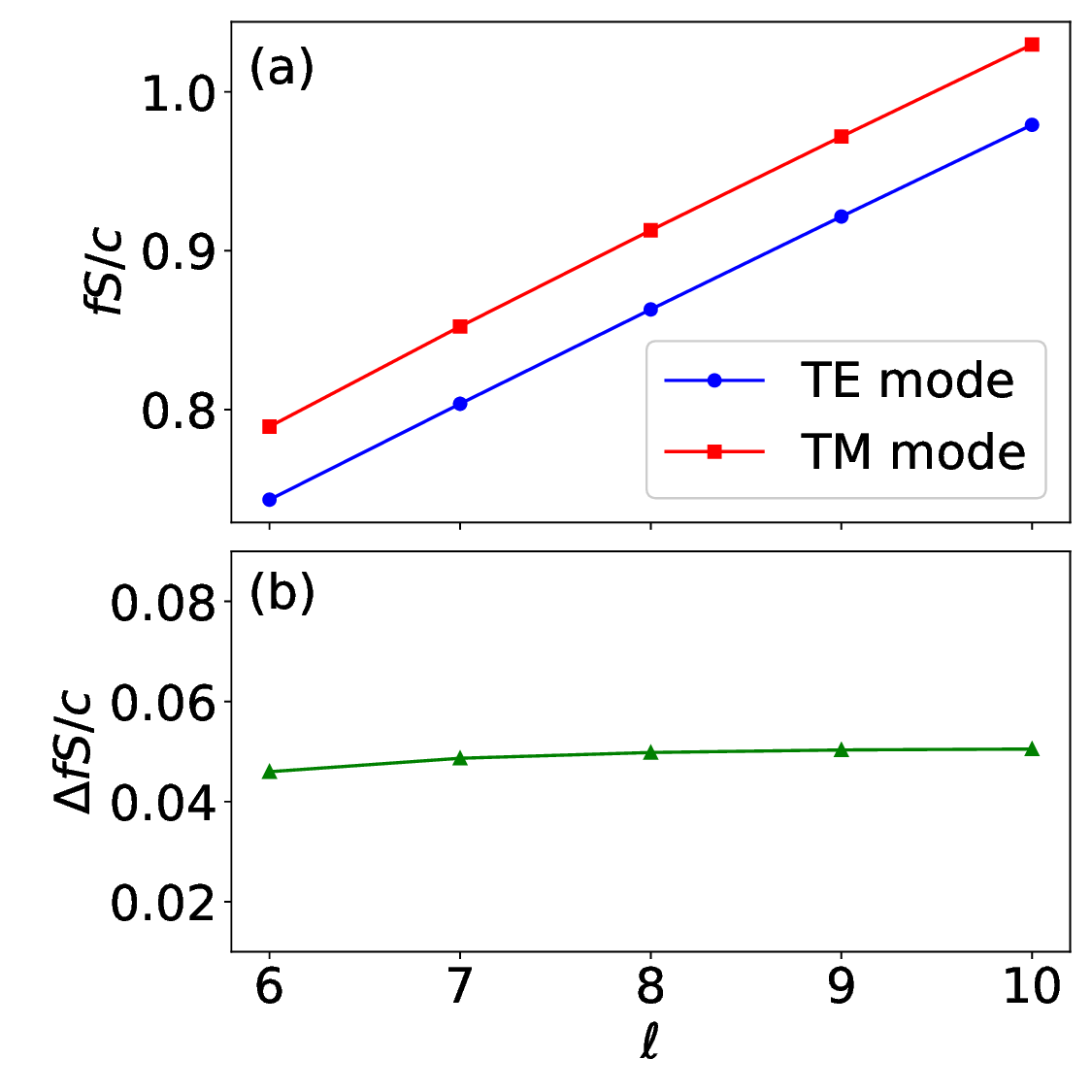}
    \caption{
    (a) Scaled resonance frequencies $( fS/c )$ for transverse electric (TE) and transverse magnetic (TM) modes of the spherical dielectric resonator under consideration as a function of the angular momentum index $ \ell $. The data highlights the splitting between TE$_{\ell}$ and TM$_{\ell}$ modes at each $\ell$. (b) The corresponding frequency difference  $\Delta fS/c = (f_{\mathrm{TM}_{\ell}} - f_{\mathrm{TE}_{\ell}})S/c$, quantifying the frequency difference as a function of $ \ell $.
    } 
    \label{fig_03}
\end{figure}

In Fig.~\ref{fig_03}(a) we show the corresponding resonance frequencies for both the TE and TM modes, in scaled units, versus the $\ell$ index, for $\ell$ ranging from $6$ to $10$. Such values of $\ell$ are typical for high-$Q$ Mie modes. We see an almost linear increase in the Mie mode frequencies with increasing $\ell$. In Fig.~\ref{fig_03}(b) we show the scaled frequency difference $\Delta fS/c$ of modes with the same $\ell$. We see that it remains almost constant $\Delta fS/c \simeq 0.05$. This means that for axion masses at the range of $\sim\mu$eV$/c^2$  the particle radius would be in the millimeter scale, while the EM radiation will be in the microwave range. Such spherical resonators exhibiting Mie modes can be fabricated and measured experimentally using microwave radiation~[\onlinecite{geffrin2012magnetic,lukyanchuk2021colossal,dey2021low,subrahmannian2022study}].  
Microwave coupling to and from the dielectric resonator can be achieved using a small loop or dipole antenna placed in the near field, enabling efficient excitation and extraction of the desired resonant modes. The output (converted) photons may be collected by a frequency- and polarization-selective antenna optimized for the corresponding mode, as commonly realized in microwave resonator experiments.
{Since the frequency separation between the initial and final photons is on the order of hundreds of MHz, well above the typical bandwidth of both the pump and the detector, no interference between the input and output signals is expected. This separation is further reinforced by the fact that the two signals also possess opposite polarization signatures.
We note also here that no external static magnetic field has to be applied in the proposed setup.}
We emphasize that the proposed platform is fully scalable and can be adjusted to the targeted axion mass. In Table~\ref{table2} we show the correspondence guidelines towards the selection of the resonator parameters for targeted axion searches.

\begin{table}[h!]
\centering
\caption{Axion Mass, Resonator Size, and Electromagnetic Frequency Correspondence.}
\label{table2}
\begin{tabular}{c|c|c} 
\hline
\textbf{Axion mass} & \textbf{Radius of the resonator} & \textbf{Frequency of EM radiation} \\
\hline
\hline
1 eV$/c^2$   & 62.5 nm     & ultra violet band          \\
\hline
1 meV$/c^2$  & 62.5 $\mu$m     & terahertz band       \\
\hline
1 ${\mu}$eV$/c^2$  & 62.5 mm     & microwave band       \\
\hline
\end{tabular}
\label{table2}
\end{table}

Furthermore, by targeting a specific mass window (e.g., an octave), it is possible to scan a portion of this window while keeping the resonator size constant and varying the refractive index of the surrounding medium instead. For example, embedding the resonator in a mixed-liquid medium, where each liquid has a different refractive index (typically $\approx1$ to $1.4$) and is transparent in the microwave range, adjusting the concentration of the liquids allows for a gradual change in the effective refractive index $n_{\rm{m}}$ of the embedding medium, permitting continuous scanning over roughly half the desired octave, since the new frequency difference scales approximately as $\Delta fSn_{\rm{m}}/c$.

Within this range of axion masses and particle radii, we can proceed with the calculation of the respective transition rates. 
The transition matrix element for axion absorption from Eq.~\ref{eq:scatelement} (making use of Eqs.~\ref{TE},\ref{TM}), after some straightforward algebra becomes

\begin{eqnarray}
    G &=& \frac{\pi g_{\alpha\gamma\gamma}}{2}   \int_V{\rm d}^{3}r A({\bf{r}}) \big(
    {\bf E}^*_{E\ell m}({\bf r};\omega_{\mathrm{f}})\cdot {\bf B}_{M\ell m}({\bf r};\omega_{\mathrm{i}}) 
    -
    {\bf B}^*_{E\ell m}({\bf r};\omega_{\mathrm{f}})\cdot {\bf E}_{M\ell m}({\bf r};\omega_{\mathrm{i}})
    \big) \nonumber \\
    &=& \frac{\pi g_{\alpha\gamma\gamma} A}{2}  
    \sqrt{\epsilon\epsilon_0\mu\mu_0}
    a^*_{\mathrm{f};E\ell m}a_{\mathrm{i};M\ell m}
    I(\ell;S,q_{\mathrm{i}},q_{\mathrm{f}}),
    \label{calcG}
\end{eqnarray}
where 
$I(\ell;S,q_{\mathrm{i}},q_{\mathrm{f}})= \int_0^S r^2 dr \big[ (\ell^2 + \ell + 1)\frac{j_{\ell}(q_\mathrm{i} r)j_{\ell}(q_\mathrm{f} r)}{q_\mathrm{i} q_\mathrm{f} r^2} +j'_{\ell}(q_\mathrm{i} r)j'_{\ell}(q_\mathrm{f} r)+j_{\ell}(q_\mathrm{i} r)j_{\ell}(q_\mathrm{f} r)
+j'_{\ell}(q_\mathrm{i} r) \frac{j_{\ell}(q_\mathrm{f} r)}{q_\mathrm{f} r}
+j'_{\ell}(q_\mathrm{f} r) \frac{j_{\ell}(q_\mathrm{i} r)}{q_\mathrm{i} r}
\big]$.
In Eq.~\ref{calcG} we assumed $A({\bf{r}})$ to be homogeneous in the volume of the sphere, i.e., $A({\bf{r}}) \approx A$, since the de Broglie wavelength of the axion is very large (at the order of kilometers~[\onlinecite{semertzidis2022axion}]). The coefficients $a_{\mathrm{i};M\ell m}$, $a_{\mathrm{f};E\ell m}$ can be calculated analytically with respect to the one-photon vacuum energy\mbox{[\onlinecite{almpanis2020spherical}]}.

Now, we proceed to estimate the order of magnitude of the photon-to-photon transition rate mediated by galactic axions with mass $m_{\alpha}=1~\mu$eV$/c^2$. For this mass, the radius of the spherical dielectric resonator must be set at $62.5$~mm so that, for $\ell=10$ the input TE$_{\ell=10}$ resonates at $4.697$~GHz ($\hbar\omega_{\rm{i}}=1.943\times10^{-5}$~eV), while the output TM$_{\ell=10}$ mode resonates at $4.939$~GHz ($\hbar\omega_{\rm{f}}=2.042\times10^{-5}$~eV). 
We set an indicative $|\tilde{g}_{\alpha\gamma\gamma}|$ at the order of {$\sim 10^{-13}$~GeV$^{-1}$. To compute the axion field amplitude $|A|$ we use the local dark matter density, typically taken to lie in the range $3\times10^{14} \lesssim \rho_{DM} \lesssim 4\times10^{14}$~eV$/$m$^3$~\mbox{[\onlinecite{marsh2019proposal,kim2019effective,roising2021axion,berlin2024absorption}]}, where we use the formula $|A|=\sqrt{2\rho_{DM}\hbar^3/c}/m_{\alpha}$ from Ref.~\mbox{[\onlinecite{kim2019effective}]}, which results in $|A|\simeq2.48\times10^3$~eV for $\rho_{DM}=4\times10^{14}$~eV$/$m$^3$.
Based on these inputs, the prefactor $\frac{\pi}{2}\frac{|\tilde{g}_{\alpha\gamma\gamma}||A|}{\mu_0 c}\sqrt{\epsilon\epsilon_0\mu\mu_0}$ evaluates approximately {$ 1.2\times10^{-29}$~A$\cdot$s/V$\cdot$m}.
The field amplitudes $|a_{\mathrm{i};M\ell m}|$, $|a_{\mathrm{i};E\ell m}|$ normalized to the one-photon vacuum energy\mbox{[\onlinecite{almpanis2020spherical}]}
($\frac{1}{4}\int_V(\epsilon\epsilon_0|E|^2+\frac{1}{\mu_0}|B|^2)=\hbar\omega_{\rm{i(f)}}$)
of the initial and the final photon state, respectively, are found to be $|a_{\mathrm{i};M\ell m}|= 4.76\times10^{-5}$~V/m and $|a_{\mathrm{f};E\ell m}|=5.10\times 10^{-5}$~V/m. The radial integral yields $|I(\ell;S,q_{\mathrm{i}},q_{\mathrm{f}})|=3.85\times10^{-7}$~m$^3$. Substituting all values into Eq.\mbox{\ref{calcG}}, we obtain a matrix element magnitude {$|G| \simeq 0.7\times10^{-25}$~eV}. 
Applying Fermi's golden rule, i.e., $R_{\alpha\gamma\gamma}=\frac{2\pi}{\hbar}|G|^2\varrho(E_{\mathrm{f}})$ and using an optical density of states $\varrho(E_{\mathrm{f}})\simeq \frac{1}{\pi}\frac{2}{h \Gamma_{\rm{f}}}\simeq 4.8\times10^8$~eV$^{-1}$, where $\Gamma_{\rm{f}}\simeq 3.2\times10^{-4}$~GHz is the full width at half maximum of the final resonance (the quality factor is $Q_{\rm{f}}\simeq\frac{\hbar\omega_{\rm{f}}}{h\Gamma_{\rm{f}}}\simeq 1.5\times 10^4$), we find a photon-to-photon transition rate
{$R_{\alpha\gamma\gamma}\simeq 0.2\times10^{-25}$~Hz}. 
In practice, dielectric resonators are pumped with multiple photons, which accumulate in the cavity and enhance the transition probability. Assuming that the maximum sustainable electric field in silicon (that better corresponds to the given relative permittivity) is $E_{max}\simeq3\times10^7$~V/m~\mbox{[\onlinecite{singh2006reliability,kim2017dielectric}]}, the corresponding theoretical limit on the number of stored photons is $N \lesssim 1.5\times 10^{25}$ ($N=\frac{U_{\rm{max}}}{\hbar\omega_{\rm{i}}}$, where $U_{\rm{max}}=\frac{1}{2}\epsilon\epsilon_0\int_V|E_{\rm{max}}|^2d\tau$ the maximum stored energy inside the resonator). This leads to a maximum 
transition rate of
{$\Tilde{R}_{\alpha\gamma\gamma}=N\times R_{\alpha\gamma\gamma} \simeq 3.4 \times 10^{-2}$~Hz}.  
{The corresponding output power (photon energy $\times$ transition rate) $P_{\rm{out}}=(\hbar\omega_{\rm{f}})\times \tilde{R}_{\alpha\gamma\gamma}$ would be 
at the order of $\sim 10^{-25}$ W, which is a theoretical upper limit for the given setup. However, in practice, the number of accumulated photons inside the cavity is determined by the input power $P_{\rm{in}}$, the quality factor of the incoming resonance frequency $Q_{\rm{i}}$ and the corresponding frequency $f_{\rm{i}}$, through the relation $N=P_{\rm{in}}Q_{\rm{i}}/2\pi h f^2_{\rm{i}}$.
For a realistic continuous input power of $P_{\rm{in}}=100$~W, the number of accumulated photons in the resonator will be $N \simeq 2\times10^{19}$, where $Q_{\rm{i}}\simeq 2\times10^4$. This results in output power at the order of $P_{\rm out}\sim10^{-30}$~W, which, although small, is not unusual of dark-matter searches. 
}

\begin{figure}[h]
    \centering
    \includegraphics[width=1.0\linewidth]{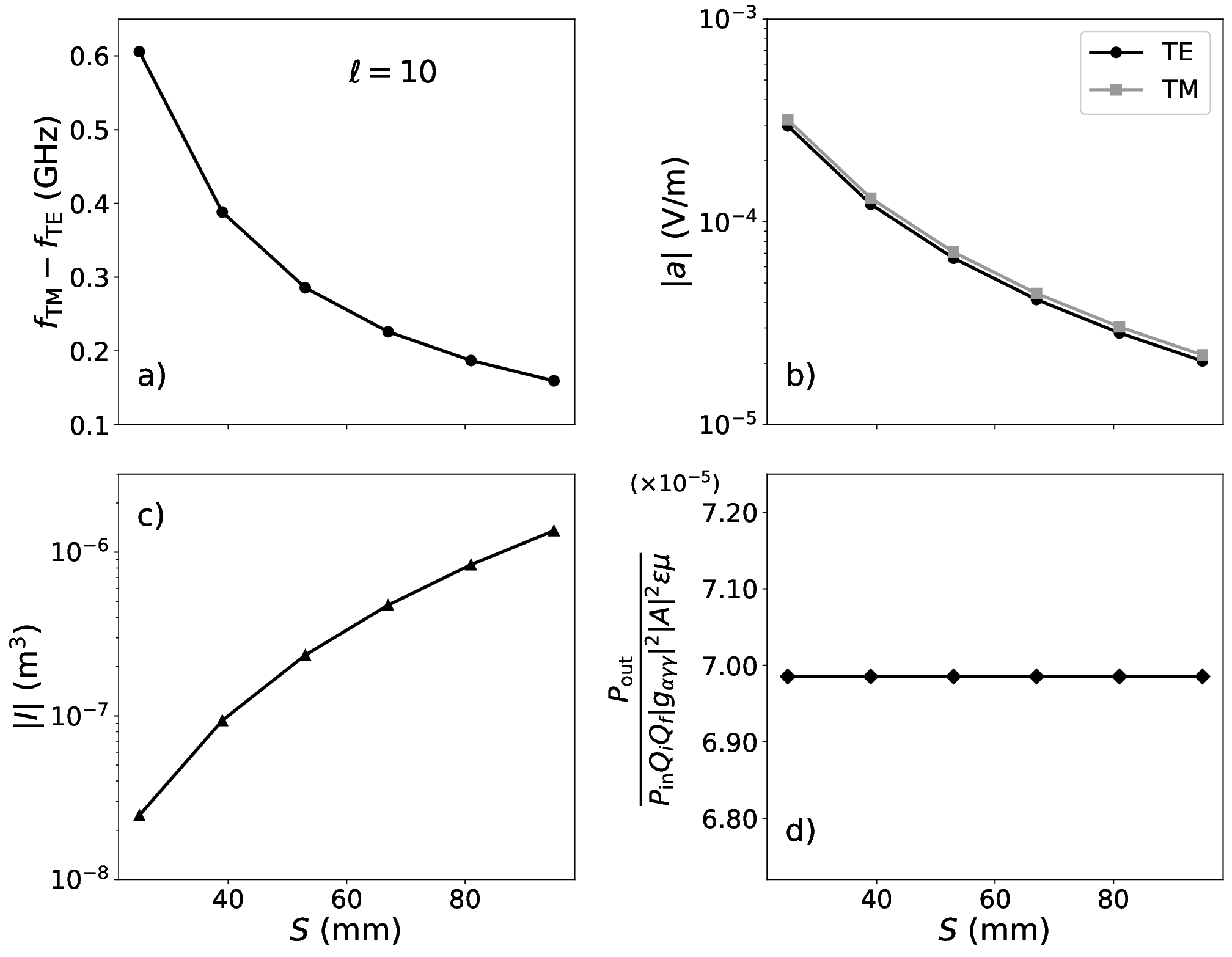}
    \caption{
 {   (a) Frequency difference between transverse magnetic (TM) and transverse electric (TE) Mie modes, $f_{\mathrm{TM}} - f_{\mathrm{TE}}$, for angular momentum $\ell = 10$ as a function of the sphere radius $S$. 
The secondary vertical axis indicates the equivalent axion mass
$m_{\alpha}$ ($\mu\mathrm{eV}/c^{2}$) corresponding to the mode
splitting.
(b) Magnitude of the corresponding modal amplitudes $|a|$ for TE and TM modes, showing comparable scaling behavior with increasing $S$. 
(c) Absolute value of the overlap integral $|I|$ entering the transition matrix element, exhibiting a monotonic increase with the resonator size. 
(d) Normalized output power 
$P_{\mathrm{out}} / \left[ P_{\mathrm{in}} Q_{\rm i} Q_{\rm f} |g_{a\gamma\gamma}|^2 |A|^2 \epsilon \mu \right]$ (dimensionless quantity), 
which remains approximately constant across the considered range of $S$.
}}
    \label{fig_04new}
\end{figure}

{To find ways for improving the output signal strength, we next investigate the scaling of $P_{\rm out}$ with the relevant intrinsic and external parameters of the system, paying particular attention to the scatterer size. 
In the upconversion approach, the first key step is to tailor the frequency splitting between the TE and TM modes so that it matches the unknown axion mass under investigation. As shown in Fig.\mbox{\ref{fig_04new}}(a), the separation $f_{\mathrm{TM}} - f_{\mathrm{TE}}$ $(= m_{\alpha} c^2 /h)$ can be continuously tuned by varying the sphere radius $S$, thereby selecting a specific target axion mass window at fixed material parameters. 
Although Fig.~\mbox{\ref{fig_04new}}(a) shows only the representative case $\ell=10$, the same inverse-radius trend ($m_{\alpha}\propto S^{-1}$) persists for higher-order Mie modes. Quantitatively, within the asymptotic description[\mbox{\onlinecite{lam1992explicit}}], the TE-TM frequency splitting at $S=25~\mathrm{mm}$ decreases from about $0.59~\mathrm{GHz}$ for $\ell=10$ to about $0.54~\mathrm{GHz}$ for $\ell=100$. At $S=95~\mathrm{mm}$, the corresponding values are approximately $0.155~\mathrm{GHz}$, and $0.142~\mathrm{GHz}$, respectively. Thus, increasing $\ell$ up to $\ell=100$ produces only a modest reduction of the TE-TM splitting, while the dominant dependence remains the inverse scaling with the sphere radius.
The other two panels of Fig.~\mbox{\ref{fig_04new}} clarify how the relevant photon-cavity quantities entering the transition matrix element $G$ evolve with $S$. Panel (b) shows that the one-photon field amplitudes of the TE and TM modes of the same $\ell$ remain of comparable magnitude and decrease similarly with increasing radius. Panel (c) shows that the radial overlap integral $|I|$, which quantifies the spatial overlap between the initial and final resonant modes in Eq.~(18), increases monotonically with $S$.
Now, we express the output power $P_{\rm{out}}$ extracted from the spherical dielectric resonator in terms of the phenomenological parameters, in order to elucidate how it scales with the relevant cavity characteristics:}

\begin{equation}
 \frac{P_{\mathrm{out}}}{P_{\mathrm{in}}Q_{\rm i}Q_{\rm f}\left|\tilde g_{a\gamma\gamma}/(\mu_0 c)\right|^2 |A|^2 \epsilon\mu} =  
    \pi^2 \frac{|a_{TM}|^2 |a_{TE}|^2 |I|^2} {(h f_{\rm{i}})^2} 
    \epsilon_0 \mu_0  
    \label{eq:P_out}
\end{equation}

{The right-hand side of Eq.~\mbox{\ref{eq:P_out}} remains nearly constant under radius variation, as shown in Fig.~\mbox{\ref{fig_04new}}(d). The weak variation observed in Fig.~\mbox{\ref{fig_04new}}(d) indicates that, for the spherical Mie modes considered here, the photon-cavity characteristics (including the quality factors, which are independent of $S$) remain practically unaffected by the size of the spherical resonator. This provides the freedom to target different axion masses by varying the radius $S$ without significantly altering the cavity-dependent contribution to the output power, apart from limitations imposed by the material breakdown threshold. We note, however, that the axion field amplitude $|A|$ is not strictly independent of $S$. Since $|A| \propto m_\alpha^{-1/2}$ and the resonant condition implies approximately $m_\alpha \propto S^{-1}$ [see Fig.~\mbox{\ref{fig_04new}}(a)], we find $|A|^2 \propto S$. Therefore, while the factor shown in Fig.~\mbox{\ref{fig_04new}}(d) remains essentially unchanged, the absolute output power acquires an additional approximately linear dependence on the resonator radius through the axion field amplitude. We now define the quantity
$ \frac{P_{\mathrm{out}}}{P_{\mathrm{in}}\left|\tilde g_{a\gamma\gamma}\right|^2 |A|^2} \equiv C(\ell)$,
which depends only on $\ell$. Mathematically, according to the asymptotic relations~[\mbox{\onlinecite{lam1992explicit}}], the parameter $C(\ell)$ grows with $\ell$ approximately as $C(\ell)\propto \ell^4$, indicating a preference for modes with higher $\ell$ due to their higher quality factors. Nevertheless, in physically realizable systems, this enhancement cannot be pursued indefinitely. Practical considerations, including material losses, finite axion coherence time, and other physical constraints, impose an upper limit on the usable cavity quality factors, as will be discussed later on.
}

To further enhance the sensitivity limit within the proposed physical framework, several advanced techniques can be explored, pushing the boundaries of current technological capabilities. Whispering gallery modes (WGMs) that are photonic resonances with high angular momentum indices $\ell$, can achieve $Q$-factors on the order of $\sim10^5$ to $10^{10}$~\mbox{[\onlinecite{gorodetsky1996ultimate,yu2012spherical,rueda2016efficient,vogt2018ultra,lambert2020coherent,ilchenko2013whispering}]}. {From a practical standpoint, in whispering‑gallery modes the excitation is naturally dominated by modes with $|m|=\ell$~\mbox{[\onlinecite{haigh2016triple,vogt2018ultra}]}, which are concentrated near the surface of the sphere, so the requirement $\delta_{mm'}$ is automatically satisfied, while the spectral separation between TE and TM WGMs with the same indices can be predicted analytically~\mbox{[\onlinecite{lam1992explicit}]} despite the richness of the spectrum.} 
{If we assume, as a moderate case, $\ell=14$, then the mathematical values of our lossless-cavity quality factors grow to $Q\sim 10^{10}$. 
}
This improvement would significantly increase the optical density of states $\varrho(E_{\mathrm{f}})$ by additional orders of magnitude. However, when the resonator quality factor exceeds the axion field's intrinsic quality factor
($Q_{\alpha}\sim10^{6}$), the effective enhancement of the optical density of states 
becomes limited by $Q_{\alpha}$, and further increments do not contribute additional enhancement of ${R}_{\alpha\gamma\gamma}$.
{By taking this into account, the mathematical value of the parameter $C(\ell)$ is enhanced by a factor $\sim10^7$.}
Furthermore, materials such as silicon carbide and fused silica or diamond that have low intrinsic losses and can handle higher electric fields than silicon by one and two orders of magnitude, respectively\mbox{[\onlinecite{singh2006reliability,denisenko2005diamond}]}, are capable of sustaining larger numbers of photons ($N\sim 10^{27}$ to $10^{29}$) without breaking down. We note however that refractive-index nonlinearities may occur at high field intensities, introducing additional noise into the measured signal. Therefore, materials such as diamond, which possess exceptionally high thresholds for nonlinear effects, may be preferable. 
{To efficiently pump more photons inside the cavity, advanced pumping technologies such as Klystrons could be explored to provide the necessary energy and maximize the number of photons $N$. As today's Klystrons can provide orders of magnitude higher input power to the resonator, than the $P_{\rm in}=100$~W assumed before, a direction for future efforts would be to mitigate overheating while increasing the input power, with strategies such as pulsed pump operation as well as the development of novel thermal management techniques, to approach the theoretical upper limit of detectability. As an optimistic example, $P_{\rm in}=10$~kW on a diamond resonator at a cryogenic environment, would increase the output power by a factor $\sim 10^2$. These two improvements, combined, could substantially increase the output power previously calculated by an order $\sim 10^{9}$, resulting in $P_{\rm out}\sim10^{-21}$~W, a level within reach of state-of-the-art cryogenic microwave receivers.} 

{The signal-to-noise ratio (SNR) in a haloscope-type search is governed by the Dicke radiometer equation}

\begin{equation}
{\rm{SNR}} = \frac{{P_{\rm{out}}}}{k_B T_{sys}} \sqrt{\frac{t_{int}}{w}},
\end{equation}
{where $k_B$ is the Boltzmann’s constant, $T_{sys}$ the total system noise temperature (including all thermal, quantum, and detector contributions), $w$ the detection bandwidth, and $t_{int}$ the integration time per frequency step. For a cryogenic receiver with $T_{\rm sys}\simeq5$~K and
$w\simeq1$~Hz, it is $P_{\rm out}/(k_B T_{\rm sys}) \simeq 14.5$. Requiring a target sensitivity of ${\rm SNR}\simeq5$ yields
$t_{\rm int}=(5/14.5)^2\simeq1.2\times10^{-1}\ {\rm s}$. Thus, an SNR of order five can be achieved with an integration time of approximately $t_{\rm int}\sim0.1$~s per frequency step. The expected axion signal bandwidth is
$\Delta f_{\alpha}\sim f_{\alpha}/Q_{\alpha}$. For an axion mass $m_{\alpha}=1~\mu{\rm eV}/c^2$, the corresponding axion frequency is $f_{\alpha}=m_{\alpha}c^2/h\simeq2.4\times10^8~{\rm Hz}$, giving $\Delta f_{\alpha}\sim2.4\times10^2~{\rm Hz}$. Choosing the scan step size to be comparable to the axion linewidth, $\Delta f_{\rm step}\simeq\Delta f_{\alpha}$, corresponds to an effective scan quality factor $Q_{\rm scan} \equiv f_{\alpha}/\Delta f_{\rm step} \sim Q_{\alpha} \sim 10^6$. The number of frequency steps required to cover one octave (a factor-of-two span in axion frequency) is therefore $N_{\rm steps} \simeq Q_{\rm scan}\ln2 \simeq 7\times10^5$. With $t_{\rm int}\simeq0.12$~s per step, the corresponding scan time is $T_{\rm oct} = N_{\rm steps}  t_{\rm int} \approx 23~{\rm h}$. Consequently, an octave in frequency can be scanned in approximately
one day under these assumptions. We should also keep in mind that different sphere sizes allow parallel searches.
Finally, we note that recent progress in superconducting microwave single-photon detectors operating at GHz frequencies~[\mbox{\onlinecite{pankratov2022towards,balembois2024cyclically,braggio2025quantum,kim2025modeling}}] suggests an alternative, fully digital readout scheme, in which the SNR is governed by photon-counting statistics rather than by an effective noise temperature. Such devices have demonstrated rapid advances in sensitivity and may eventually provide a complementary route toward the detection of extremely weak axion-induced signals. Nevertheless, their deployment in a broadband, long-term axion haloscope search remains technologically more challenging than the cryogenic linear-receiver approach adopted here.
}

{Moving to higher axion masses, for instance to
$m_{\alpha}=1~\mathrm{meV}/c^2$, the characteristic frequency of the
signal is pushed into the terahertz regime and the effective mode
volume of the dielectric resonator is correspondingly reduced
($S=62.5~\mu\mathrm{m}$ - see Table~\mbox{\ref{table2}}).
In this frequency range the situation is qualitatively different
from the microwave case discussed above: single-photon detection
technology in the infrared and THz domains is substantially more
mature, offering high quantum efficiencies and low
dark-count rates. It is therefore natural, at these higher axion
masses, to abandon the purely radiometric readout paradigm and to
consider instead a photon-counting detection scheme. We note, however, that reducing the resonator radius to
$S=62.5~\mu\mathrm{m}$ decreases $|A|^2$ by a factor of $10^{-3}$,
as discussed earlier.
Since $P_{\rm out}\propto
|\tilde g_{\alpha\gamma\gamma}|^2 |A|^2$, the same output power
can be retained by increasing the coupling constant $|\tilde{g}_{\alpha\gamma\gamma}|$ by a factor
$\sqrt{10^3}$. Thus, instead of
$|\tilde g_{\alpha\gamma\gamma}|=10^{-14}~\mathrm{GeV}^{-1}$, we use
$|\tilde g_{\alpha\gamma\gamma}|\simeq
3.16\times10^{-13}~\mathrm{GeV}^{-1}$ for this higher-mass reference
case. In this estimate, the previously assumed
input power should be understood as an optimistic
peak pulsed power, rather than as a continuous-wave input power for a
micron-scale dielectric resonator. For a single-photon detector with
overall efficiency $\eta$ and dark-count rate $R_{\rm dark}$, the
number of detected signal photons in an integration time $t_{\rm int}$
is
$N_{\rm sig}=\eta P_{\rm out}t_{\rm int}/(h f_0)$, while the
background contribution is
$N_{\rm dark}=R_{\rm dark}t_{\rm int}$. The corresponding
signal-to-noise ratio is}

\begin{equation}
{\rm SNR} =
\frac{N_{\rm sig}}{\sqrt{N_{\rm sig}+N_{\rm dark}}}\,.
\end{equation}

{In the ideal shot-noise limit ($N_{\rm sig}\gg N_{\rm dark}$),
we obtain
$t_{\rm int}^{(\mathrm{shot})}
= {\rm SNR}^2 h f_0/(\eta P_{\rm out})$.
For representative values
$f_0\simeq5~\mathrm{THz}$, $\eta\simeq0.5$,
$P_{\rm out}=10^{-21}~\mathrm{W}$, and a target
${\rm SNR}\simeq5$, this yields
$t_{\rm int}^{(\mathrm{shot})}\sim170~\mathrm{s}$ per frequency step.
For an axion mass $m_{\alpha}=1~{\rm meV}/c^2$, the corresponding axion frequency is $f_{\alpha}=m_{\alpha}c^2/h\simeq2.4\times10^{11}~{\rm Hz}$, giving $\Delta f_{\alpha}\sim2.4\times10^5~{\rm Hz}$.
Choosing the scan step size
$\Delta f_{\rm step}\simeq\Delta f_{\alpha}$ therefore corresponds
to an effective scan quality factor
$Q_{\rm scan}\equiv f_{\alpha}/\Delta f_{\rm step}\sim Q_{\alpha}\sim10^6$.
The number of independent frequency steps required to cover one octave
is then
$N_{\rm steps}\simeq Q_{\rm scan}\ln2\approx7\times10^5$, and the
total scan time
$
T_{\rm oct}= N_{\rm steps}\,
t_{\rm int}^{(\mathrm{shot})} 
\approx3.7~\mathrm{years}.
$
Including realistic dark counts in the dark-count-limited regime modifies the required integration
time according to
$
t_{\rm int}^{(\mathrm{dark})}
=
\left(
{\rm SNR}\,h f_0/(\eta P_{\rm out})
\right)^2
R_{\rm dark},
$
which can lengthen $T_{\rm oct}$ further, depending on the achievable
dark-count rate of the infrared/THz detector. However, parallel operation of multiple resonators could offer a path toward substantially
accelerating the overall survey time.} {We also note that the proposed scheme achieves its maximum sensitivity under triply resonant conditions, with the enhancement progressively diminishing as the detuning between the axion frequency and the cavity mode splitting exceeds the effective linewidth.}

{The above discussions correspond to an optimistic scenario
illustrating the ultimate potential of the proposed framework.
However, a more moderate and technologically conservative benchmark would be also instructive.  
Specifically, we now assume a similar implementation driven with a
continuous input power of $P_{\rm in}=100~{\rm W}$ and include a
representative damping of the signal due to thermal
losses in the resonator by assuming an effective quality factor $Q_{\rm eff}\simeq10^{-1}Q$ for each resonant mode. Although such losses are expected to remain
relatively small in low-loss dielectric materials operated at
cryogenic temperatures, they reduce the achievable output power
compared to the optimistic estimate. 
To facilitate a direct comparison with the optimistic scenario, we
retain the same target sensitivity (${ \rm SNR}=5$), integration
times, scan rates and octave-scan durations. Under these assumptions,
the corresponding couplings become
$|\tilde g_{a\gamma\gamma}| \simeq 10^{-11}\,{\rm GeV}^{-1}$ at
$m_a=1~\mu{\rm eV}/c^2$
and
$|\tilde g_{a\gamma\gamma}| \simeq 3.16\times10^{-10}\,{\rm GeV}^{-1}$ at
$m_a=1~{\rm meV}/c^2$.
These values define a baseline scenario 
which may be regarded as a realistic
first-generation implementation of a respective experiment, whereas the
optimistic scenario marks the sensitivity that could become
accessible through future improvements in resonator performance,
thermal management and drive technology.}

\begin{figure}[!t]
    \centering
    \includegraphics[width=0.75\linewidth]{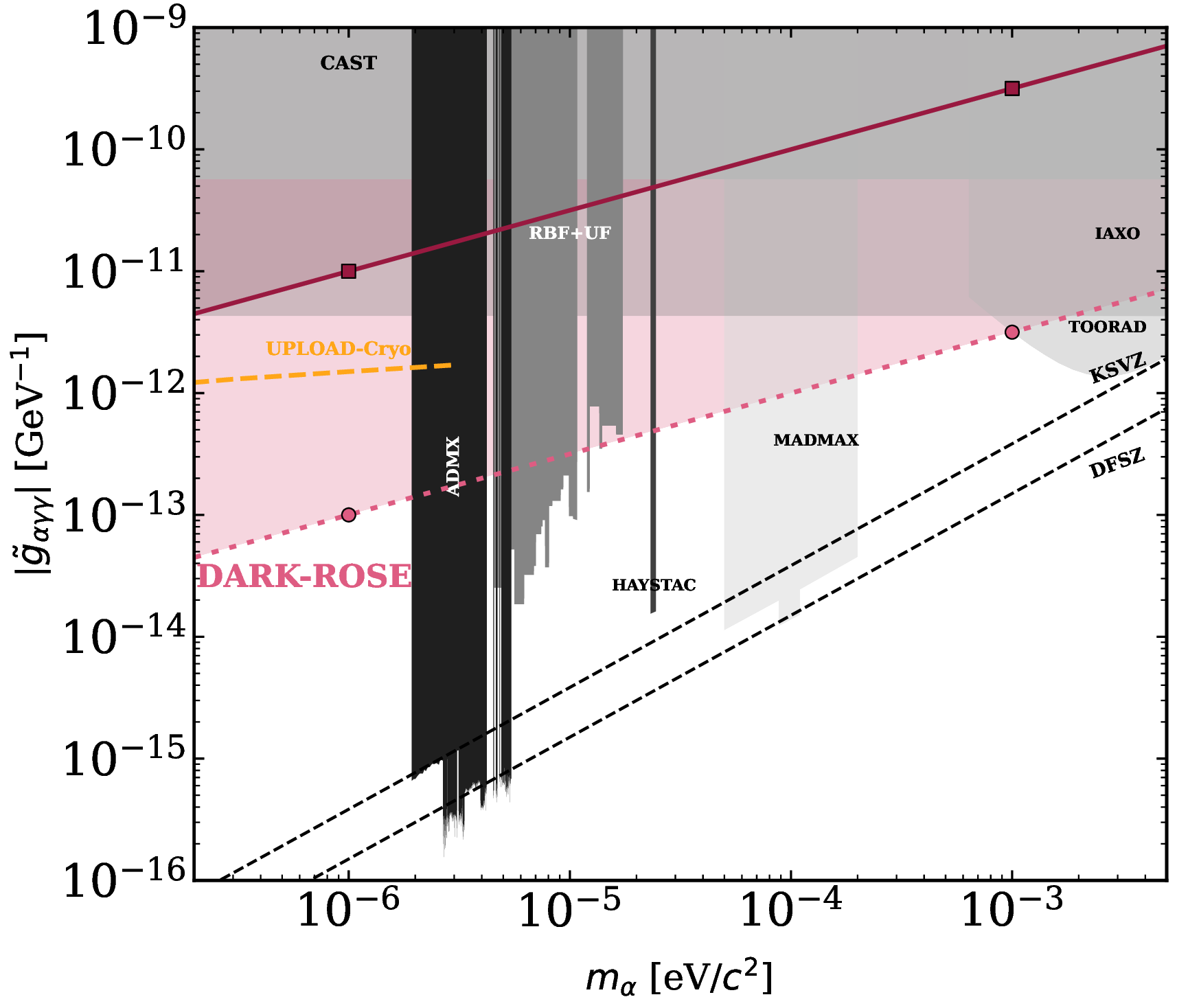}
 \caption{Constraints and projected sensitivities on the axion-photon-photon coupling
$|\tilde{g}_{\alpha\gamma\gamma}|$ as a function of the axion mass $m_a$.
{The DARK-ROSE projections are obtained from the two benchmark
calculations discussed in Sec.~V
($m_{\alpha}=1~\mu{\rm eV}/c^2$ and
$m_{\alpha}=1~{\rm meV}/c^2$),
assuming ${\rm SNR}=5$.
The dotted rose boundary corresponds to the optimistic scenario,
while the solid dark-red boundary corresponds to a baseline scenario
assuming a cryogenic $100~{\rm W}$ implementation with representative losses included.
For both scenarios, the benchmark sensitivities are defined using
integration times
$t_{\rm int}\approx0.12~{\rm s}$ (for $m_{\alpha}=1~\mu{\rm eV}/c^2$) 
and
$t_{\rm int}\approx170~{\rm s}$ (for $m_{\alpha}=1~{\rm meV}/c^2$) per frequency step, 
for an axion coherence
quality factor $Q_\alpha\simeq10^6$.
The intermediate and extrapolated DARK-ROSE sensitivities are obtained
from the scaling relations derived in Sec.~V.
The solid purple curve indicates the projected reach of the UPLOAD-II-Cryo
upconversion experiment.}
Shaded regions correspond to existing exclusion limits from CAST, RBF+UF, and ADMX,
while semi-transparent grayscale regions indicate the projected sensitivities of
HAYSTAC, TOORAD, MADMAX, and IAXO (datasets obtained from the AxionLimits repository maintained by C.~O'Hare).
The dashed lines show the benchmark KSVZ and DFSZ QCD axion models.
Both axes are logarithmic; the displayed parameter space covers
$2\times10^{-7}\le m_a \le 5\times10^{-3}~\mathrm{eV}/c^2$ and
$10^{-16}\le |\tilde{g}_{a\gamma\gamma}| \le 10^{-9}~\mathrm{GeV}^{-1}$.}
    \label{fig_05}
\end{figure}

A future experiment based on this framework could be named
DARK-ROSE, which stands for \emph{DARK matter search via Resonant
Optical Scattering Experiment}. Figure~\mbox{\ref{fig_05}}
illustrates the projected DARK-ROSE sensitivity within the broader
axion landscape. 
Existing exclusions from haloscopes (RBF+UF\mbox{[\onlinecite{depanfilis1987limits,hagmann1990results}]}, ADMX\mbox{[\onlinecite{nitta2023search,mcallister2024tunable,carosi2025search}]})
and the helioscope CAST\mbox{[\onlinecite{zioutas2005first,arik2009probing}]} delimit the currently probed parameter space, while the semi-transparent grayscale overlays indicate  projected sensitivities from HAYSTAC\mbox{[\onlinecite{zhong2018results}]} and TOORAD\mbox{[\onlinecite{marsh2019proposal,schutte2021axion}]} (haloscopes), MADMAX\mbox{[\onlinecite{beurthey2020madmax}]} (dielectric haloscope), IAXO\mbox{[\onlinecite{armengaud2014conceptual,armengaud2019physics}]} (next-generation helioscope)
{and the UPLOAD-CMC-II-Cryo upconversion experiment~\mbox{[\onlinecite{thomson2021upconversion}]}},
providing complementary (future) coverage across the $\mu$eV$/c^2$–meV$/c^2$ mass window.
{The rose-shaded DARK-ROSE region is obtained from the two benchmark
calculations discussed above,
namely for 
$m_a=1~\mu{\rm eV}/c^2$ 
and for 
$m_a=1~{\rm meV}/c^2$,
assuming a target sensitivity of ${\rm SNR}=5$.
The corresponding dotted boundary in Fig.~\mbox{\ref{fig_05}}
represents the optimistic scenario based on cryogenic operation,
high-$Q$ resonators and peak excitation. The benchmark points defining the optimistic scenario are
$(m_a=1~\mu{\rm eV}/c^2,
|\tilde g_{a\gamma\gamma}|\simeq10^{-14}~{\rm GeV}^{-1})$
and
$(m_a=1~{\rm meV}/c^2,
|\tilde g_{a\gamma\gamma}|\simeq3.16\times10^{-13}~{\rm GeV}^{-1})$.
The solid boundary denotes the baseline scenario,
corresponding to a cryogenic $100~{\rm W}$ implementation including
representative losses.
The benchmark points defining the baseline scenario are
$(m_a=1~\mu{\rm eV}/c^2,
|\tilde g_{a\gamma\gamma}|\simeq10^{-11}~{\rm GeV}^{-1})$
and
$(m_a=1~{\rm meV}/c^2,
|\tilde g_{a\gamma\gamma}|\simeq3.16\times10^{-10}~{\rm GeV}^{-1})$.
Both scenarios assume identical integration times and scan rates,
with the change in sensitivity arising solely from the adopted
experimental parameters.
The intermediate sensitivities are obtained from the scaling relations
derived in this section.
As shown in Fig.~\mbox{\ref{fig_05}}, DARK-ROSE probes a broad
axion-like parameter space extending from the sub-$\mu$eV region
towards the meV scale, overlapping part of the parameter space
addressed by resonant upconversion searches while remaining
complementary to conventional haloscope and helioscope techniques.
Although the projected sensitivity does not reach the
KSVZ and DFSZ QCD-axion bands, a future development could explore a region that is currently largely
unconstrained and demonstrates the potential of resonantly enhanced
axion-mediated photon-to-photon transitions as a novel search
strategy.}
Additionally, our approach might assist the \emph{shining through walls}~[\onlinecite{povey2010microwave,redondo2011light,gninenko2014search,mendoncca2020axion,Spector2023}] experiments, where strong laboratory-produced electromagnetic fields could generate axions (through the Primmakoff effect) that then pass through opaque barriers and be detected on the far side. Apart from the (dark matter) axions considered in this work, the underlying mechanism and theoretical framework may also be extended to encompass solid-state axions (axion quasiparticles). In particular, if the resonator material is chosen to be an antiferromagnetically-doped topological insulator~[\onlinecite{marsh2019proposal,schutte2021axion,esposito2023optimal}] or a bianisotropic~[\onlinecite{serdyukov2001electromagnetics,christofi2018metal,koufidis2024electromagnetic}]/Tellegen metamaterial~[\onlinecite{prudencio2014geometrical,shaposhnikov2023emergent,safaei2024optical,yang2025gigantic,seidov2025unbounded,jazi2025realization}], it can intrinsically support axion-like electromagnetic responses.

{Finally, it is interesting to briefly comment on the potential of the spherical dielectric system proposed here, compared to the cylindrical metallic cavity used in the UPLOAD experiment~\mbox{[\onlinecite{thomson2023searching}]}, which is a mature technology. The UPLOAD experiment is an AC axion haloscope based on a cylindrical metallic cavity, where the axion-mediated photon upconversion is expected through different cylindrical eigenmodes. The sensitivity of the setup is determined by the so-called overlap functions $\xi_{ij}=\frac{1}{V}\int {\bf{e}}_i({\bf{r}}) \cdot {\bf{b}}_j({\bf{r}}) dV$, where ${\bf{e}}_{i,j}$ and ${\bf{b}}_{i,j}$ are the mode electric and magnetic field real unit vectors, such that $0 \le \xi_{ij} \le 1$. For the cylindrical case, one finds ${\xi}_{i,j} \simeq 0.4 - 0.5$ for an optimum mode pair selection (e.g. TE$_{011}$-TM$_{020}$)~\mbox{[\onlinecite{thomson2023searching}]}. In contrast, for the spherical dielectric resonator considered here,
the normalized electric and magnetic mode profiles of TE and TM multipoles of the same $\ell$ are everywhere parallel inside the resonator volume, as can be seen from Eqs.\mbox{(\ref{TE})-(\ref{TM})}.
In this case, the corresponding overlap reaches $\xi_{ij}=1$ for any pair allowed by the selection rule. 
We emphasize that the overlap coefficients $\xi_{ij}$ are defined in terms of normalized mode profiles and therefore quantify, in the present spherical geometry, only the angular overlap between the participating modes, while the radial dependence is entirely contained in the radial overlap integral $I(\ell;S;q_i,q_f)$.
On the other hand, the matrix element $G$ (Eq.\mbox{\ref{calcG}}) for the spherical dielectric cavity  is reduced by a factor $1/\sqrt{\epsilon}$ compared to a perfectly electric conducting (PEC) spherical cavity of the same radius. The respective eigenmodes of the PEC spherical cavity have the same structure as Eqs.~\mbox{(\ref{TE})-(\ref{TM})} but without the $\sqrt{\epsilon\mu}$ factor, and their discrete wavenumbers are fixed by the PEC boundary conditions~\mbox{[\onlinecite{jackson2021classical}]}. When the fields are normalized to the one-photon vacuum energy, the photon field amplitudes in the dielectric sphere acquire an additional factor $1/\sqrt{\epsilon}$ with respect to the PEC ones. These prefactors combined are responsible for the net $1/\sqrt{\epsilon}$ reduction of the matrix element $G$ of the dielectric sphere compared to the PEC one. Overall, the spherical dielectric cavity offers maximal geometric overlap ($\xi_{ij}=1$) but a reduced matrix element $G \propto 1 /\sqrt{\epsilon}$ with respect to a PEC spherical cavity of the same volume, whereas the cylindrical metallic UPLOAD cavity has smaller overlap ($\xi_{ij}\sim0.5$) but benefits from the larger photon field amplitudes of a PEC resonator.}

\section{Conclusions}

To summarize, we investigated (dark matter) axion-mediated photonic transitions within a spherical, high-finesse dielectric resonator and derived the corresponding selection rule based on group theory. Using the first-order Born approximation, we analytically computed the enhancement of the transition rates associated with axion absorption and emission processes. 
The analytic scalings identify experimentally accessible regimes, including mm-scale spheres at microwave frequencies, and suggest scan strategies via the resonator size and the embedding medium refractive index. Importantly, the scheme operates without an external magnetic field.
These findings could have direct implications for the design of resonant axion devices. Moreover, the underlying framework may also inspire future applications in engineered axion-like quasiparticle dynamics in electromagnetic metamaterials.

\section{Data Availability Statement}

All results were obtained through analytical calculations and can be independently reproduced.

\end{document}